# Cooperative Route Guidance and Flow Control for Mixed Road Networks Comprising Expressway and Arterial Network

Yunran Di, Haotian Shi*, Weihua Zhang, Heng Ding*, Xiaoyan Zheng, Bin Ran

*Abstract*—Facing the congestion challenges of mixed road networks comprising expressways and arterial road networks, traditional control solutions fall short. To effectively alleviate traffic congestion in mixed road networks, it is crucial to clear the interaction between expressways and arterial networks and achieve orderly coordination between them. This study employs the multi-class cell transmission model (CTM) combined with the macroscopic fundamental diagram (MFD) to model the traffic dynamics of expressway systems and arterial subregions, enabling vehicle path tracking across these two systems. Consequently, a comprehensive traffic transmission model suitable for mixed road networks has been integrated. Utilizing the SUMO software, a simulation platform for the mixed road network is established, and the average trip lengths within the model have been calibrated. Based on the proposed traffic model, this study constructs a route guidance model for mixed road networks and develops an integrated model predictive control (MPC) strategy that merges route guidance, perimeter control, and ramp metering to address the challenges of mixed road networks' traffic flow control. A case study of a scenario in which a bidirectional expressway connects two subregions is conducted, and the results validate the effectiveness of the proposed cooperative guidance and control (CGC) method in reducing overall congestion in mixed road networks.

## I. INTRODUCTION

Expressways facilitate high-capacity transport, long-distance travel, and rapid traffic flow, enhancing accessibility and mobility between urban regions. Consequently, expressways have increasingly become a vital element of urban infrastructure networks. Compared to conventional arterial road networks, integrating expressways within these networks diversifies travel modes and options. This mixed road network, combining arterial roads and expressways, offers a more robust and flexible transportation system. Increasingly, such networks are becoming prevalent in urban areas [1]. However, with increasing traffic demand, arterial road networks and expressways in the mixed road network face the congestion risk, bringing economic losses and air pollution to cities. Therefore, there is an urgent need for effective traffic control strategies to alleviate congestion in mixed road networks.

Expressways and arterial roads, as distinct components of transportation systems, exhibit fundamentally different traffic flow dynamics and capacities. Consequently, most traffic optimization methods apply independent control for the expressways and arterial networks. Regarding the expressways, most studies rely on link-based traffic prediction models, such as the cell transmission model (CTM), to establish corresponding control methods. The main flow control strategies for expressway systems include variable speed limits and ramp metering. The variable speed limit is the most commonly used method of mainline flow control; it provides drivers with recommended driving speeds by displaying variable message signs through roadside infrastructure [2]-[4]. However, it may be limited by inconvenient transmission of speed limit signals and incomplete compliance with the speed limit by drivers. Ramp metering strategies rely on signal control located at expressway on-ramps to regulate traffic flow entering the expressway, thereby relieving bottleneck congestion and reducing the risk of collisions [5]-[6]. The feedback controller ALINEA [7] and its improved methods are widely used for ramp metering [8]-[9]. In coordinated strategies, the control law for multiple on-ramps is determined based on the traffic conditions in multiple areas, including several on-ramps and sections on the freeway [10].

Regarding arterial networks, scholars have proposed many traffic management and control strategies based on the macroscopic fundamental diagram (MFD). MFD describes a homogeneous road network's unimodal, low-scattering relationship between accumulation (veh) and trip completion flow (veh/s). As a region-based traffic model, MFD has emerged to improve the overall understanding of network traffic characteristics, and its insensitivity to OD makes MFD an advantageous tool for assessing the traffic state in dynamic environments. Among the various MFD-based studies, perimeter control [11]-[16] and route guidance [17]-[20] are representative efforts. MFD-based perimeter control aims to improve networkwide traffic performance by controlling the flow rates between adjacent regions' boundaries, and feedback

Submission date: April 25, 2024. This research was financially supported by the National Natural Science Foundation of China (Grant Nos. 52072108 and 51878236), the Municipal Natural Science Foundation of Hefei (Grant No. 2022020) and China Scholarship Council Awards. (*Corresponding authors: Haotian Shi, Heng Ding).

Yunran Di is with College of Civil Engineering, Hefei University of Technology, Hefei 230009, China. (email: diyunran@mail.hfut.edu.cn).

Weihua Zhang, Heng Ding and Xiaoyan Zheng are with School of Automotive and Traffic Engineering, Hefei University of Technology, Hefei 230009, China. (email: weihuazhang@hfut.edu.cn; dingheng@hfut.edu.cn; zhengxiaoyan@hfut.edu.cn).

Haotian Shi and Bin Ran are with Department Civil and Environmental Engineering, University of Wisconsin-Madison, WI 53706, USA. (email: hshi84@wisc.edu; bran@wisc.edu).

control and model predictive control (MPC) methods are widely used in perimeter control. MFD-based route guidance concerns how to distribute demand flows to each route. Route guidance mostly adopts the route choice principle of dynamic system optimal to minimize the total network travel time or maximize the network outflow [21].

However, the research methodologies described above for the two systems independently do not apply to a mixed road network scenario. The arterial and expressway can exchange traffic flows via ramps for a mixed road network, so there is inherent coupling and competition between the two systems. For example, ramp metering strategies help keep the expressway system operating efficiently but may create queues at on-ramps that can propagate and clog urban centers. In the case of heavy congestion in urban networks, the continuous traffic demand from expressways, the limited space on off-ramps, and connected downstream intersections in arterial networks will escalate the existing congestion, leading to spillover to the mainstream of the expressway. In addition, the traffic status of either system in a mixed road network affects vehicle routing, and unreasonable route guidance measures may concentrate traffic on the same traffic system and cause congestion. Therefore, the harmonization of the two systems should be considered.

Most existing studies on synergistic arterial networks and expressways are limited to coordination at the local scale, including control strategies for on-ramps and adjacent intersections [22]-[23] and control strategies for off-ramps and adjacent intersections [24]-[25]. These studies focus on improving local traffic efficiency by synergistically optimizing ramp metering and intersection signal timing schemes. However, the large scale of urban road networks and the rapid propagation of congestion make local control ineffective in alleviating overall congestion. There is limited research on macro-coordinated control of arterial networks and expressways. The methods in [26]-[27] focus on modeling and controlling urban intersections and road segments, which, due to computational constraints, are more suitable for localized mixed road networks. Additionally, the coordinated control scheme combining perimeter control within subregions and ramp metering has been proposed in [1] and [28], but the positive role of traffic guidance in solving traffic congestion has not been considered. Compared to conventional arterial networks, the primary distinction of mixed road networks lies in including expressways, which provide alternative travel routes. While [29] discusses route guidance, it overlooks traffic assignment in mixed road networks. The study assumes vehicles randomly select travel routes without guidance, disregarding the influence of traffic states on route selection between expressways and arterial road networks. As a result, the control outcomes struggle to accurately represent the role of route guidance strategies in cooperative control.

Based on the discussions above, to the best of our knowledge, the topic of traffic modeling and coordinated control methods for mixed road networks has not been sufficiently investigated. Regarding modeling, CTM and MFD provide beneficial tools for describing the traffic dynamics of expressways and arterial networks. However, integrating them to construct an accurate traffic model for mixed road networks still presents challenges. Existing link-based modeling methods [28]-[29] have not yet addressed tracking vehicle paths within the expressway system of mixed road networks. This leads to an inability to accurately determine whether vehicles have exited the expressway, rendering traffic flow predictions unreliable and consequently affecting the effectiveness of control strategies. Moreover, accumulation-based MFD models typically assume a constant average trip length for each region. In mixed road networks, the traffic demand on expressways and arterial networks completes their trips through different travel modes. The disparity in trip lengths between these two travel modes can influence the MFD, a factor that has not yet been addressed. Regarding collaborative control, there is a lack of effective research investigating the impact of route guidance on mixed road networks. The implementation of route guidance relies on predicting route choice behavior, which then informs the formulation of guidance schemes according to travelers' responses. In mixed road networks, travel paths are complex, requiring consideration of the states on both expressways and arterial networks.

To address these gaps, this study develops a specialized traffic model and route guidance system for mixed road networks coupled with a synergistic control strategy that integrates route guidance, perimeter control, and ramp metering. To refine the accuracy of traffic forecasts, the study incorporates a multi-class CTM tailored for the expressway system, which categorizes vehicles according to their designated routes. This approach is synergized with the MFD to craft a comprehensive traffic model that captures the complex dynamics of mixed road networks. Furthermore, this study adjusts the MFD better to reflect the varying average trip lengths with different routes, utilizing empirical data to calibrate these parameters for a mixed road network environment. The formulation of route guidance strategies takes into account route choice behaviors. By applying the newly developed traffic model, the study computes and evaluates travel times across different routes, establishing a route choice model founded on the principles of stochastic user equilibrium. These insights guide the development of route guidance strategies, strategically influencing travel decisions to enhance traffic distribution and alleviate congestion within mixed road networks.

The main contributions of this study are as follows:

(1) Multi-class CTM and MFD are employed to model expressways and arterial subregions. This enables effective tracking of vehicle routes within the mixed road network, thus allowing for the development of an accurate traffic prediction model.

(2) Considering the route choice behavior of vehicles, a route guidance strategy is proposed for the mixed road network. Subsequently, a cooperative control method integrating route guidance, ramp metering, and perimeter control is developed, effectively relieving traffic congestion on the mixed road network.

The remainder of this study is organized as follows. Section

II is the problem description. Section III provides the traffic flow modeling approach for the mixed road network. Section IV presents the route guidance model for the mixed road network. Section V proposes a cooperative method of flow control and route guidance for the mixed road network. Section VI presents the case studies, and Section VII is the conclusion.

## II. PROBLEM DESCRIPTION

### A. Scenario Description and Assumptions

Consider a mixed road network, as shown in Fig. 1, which comprises two adjacent arterial subregions linked by a bi-directional expressway. The expressway serves as an alternative route for traffic between subregions and accommodates trips that originate and terminate outside these subregions. The traffic flows in opposite directions on an expressway do not interact directly. To distinguish between these two directions of the expressway, denote the expressway traveling from subregion $i$ to subregion $j$ by $E_{ij}, i \neq j$. Expressway $E_{ij}$ have an on-ramp and off-ramp in subregion $i$ and an on-ramp and off-ramp in subregion $j$, with the off-ramp positioned upstream of the on-ramp within the same subregion.

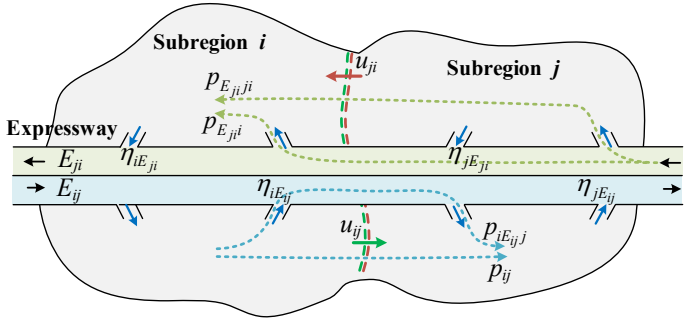

**Fig. 1.** A mixed road network with two subregions and an expressway.

To simplify the traffic flow transmission paths, the following assumptions are made for the trip routes in the mixed road network: 1) the expressway is used no more than once during each trip, i.e., a vehicle is not allowed to exit the expressway, and then re-enter it; 2) a subregion may experience a maximum of one interchange during each trip, e.g., a vehicle cannot travel from subregion $i$ to subregion $j$ and then return to subregion $i$. Given these assumptions, each trip within the mixed road network can have a maximum of two potential routes. Travelers can reach their destination directly via the arterial network or the expressway only, and those who have the option can also choose to use both the arterial network and the expressway in a single trip. The option of using expressways is available for interregional ODs; only the arterial network is used for intraregional trips. In the mixed road network shown in Fig. 1, the possible routes between each OD are shown in TABLE I. Denote the set of all possible routes in the mixed road networks as $Y$ and the set of OD pairs with more than one alternative route as $\Phi$. Any route $y \in Y$ consists of a sequence of subregions and expressway nodes.

| Origin | Destination | Routes |
|--------|-------------|--------|
| $i$ | $i$ | $y_{ii} : i \rightarrow i$ |
| $i$ | $j$ | $y_{ij} : i \rightarrow j$ |
| | | $y_{iE_{ij}j} : i \rightarrow E_{ij} \rightarrow j$ |
| $i$ | $E_{ij}$ | $y_{iE_{ij}} : i \rightarrow E_{ij}$ |
| | | $y_{ijE_{ij}} : i \rightarrow j \rightarrow E_{ij}$ |
| $i$ | $E_{ji}$ | $y_{iE_{ji}} : i \rightarrow E_{ji}$ |
| $E_{ij}$ | $i$ | $y_{E_{ij}i} : E_{ij} \rightarrow i$ |
| $E_{ij}$ | $j$ | $y_{E_{ij}j} : E_{ij} \rightarrow j$ |
| | | $y_{E_{ij}ij} : E_{ij} \rightarrow i \rightarrow j$ |
| $E_{ij}$ | $E_{ij}$ | $y_{E_{ij}E_{ij}} : E_{ij} \rightarrow E_{ij}$ |

### B. Research Framework

This study proposes a cooperative route guidance and flow control method to address the traffic congestion issue in the mixed road network. Regarding route guidance, the guidance controller allocates the traffic demand of each OD pair at a macro level to optimize the flow distribution. Regarding traffic control, perimeter controllers are placed at the boundary of the two subregions to regulate the external transfer demand. Ramp metering controllers are placed at the on-ramps to regulate the traffic flows from the subregion to the connected expressway.

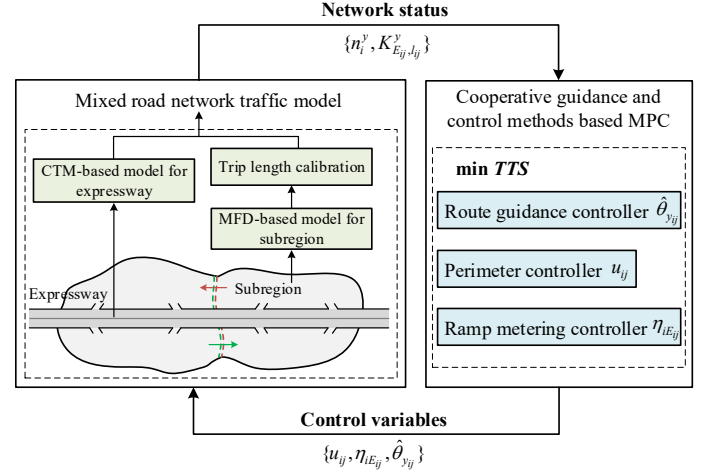

**Fig. 2.** The overall framework of cooperative route guidance and flow control.

The overall framework of cooperative route guidance and flow control is shown in Fig. 2. The study consists of two main modules: 1) modeling module, the development of a traffic flow model for the mixed road network, and 2) strategy formulation module, the formulation of cooperative guidance and control strategies. In the modeling module, this study describes the traffic dynamics of the expressway system and the subregions separately and integrates them into a mixed road network traffic model. The proposed traffic model provides a method for observing the network state, laying the foundation for formulating control strategies. In the strategy formulation module, based on the traffic state prediction in the mixed road network, a cooperative control model is established within the MPC framework to minimize the total time spent (TTS) of vehicles. The cooperative control module can generate optimal guidance and control schemes for the target road network. The details of the methodology are introduced in the subsequent sections.

## III. MIXED ROAD NETWORK TRAFFIC MODEL

This section discusses the traffic modeling module for the mixed road network, including the traffic modeling for the expressway and arterial subregions. Firstly, a multi-class CTM-based model for the expressway system is presented. Then, the MFD-based model for the arterial subregion is described. Finally, the average trip length parameter of the MFD for the mixed road network is calibrated using experimental data. The function of the mixed road network traffic model is to predict the future trend of the road network state based on the current road network state and traffic demand.

### A. CTM-based model for expressway

CTM is a commonly used link-based traffic model. This study employs CTM to model the expressway system to predict the traffic density, outflow, and average speed on the mainline and the ramps.

In the mixed road network scenario, the cell division diagram of the expressways is shown in Fig. 3. Based on the CTM theory, the expressway in each travel direction is divided into $L$ cells of equal length $L_s$. Each mainline cell contains at most one on-ramp or off-ramp. Each on-ramp and off-ramp is represented by a separate cell of length $L_s$. Assuming that the expressway is homogeneous, all cells have known triangular fundamental diagrams. For the mainline cells, the free flow speed is $V_{f1}$, the critical density is $K_{c1}$, the capacity is $C_1$, the blockage density is $K_{j1}$, and the congestion wave speed is $\omega_1$; for the ramp cells (including the on-ramps and off-ramps), the free flow speed is $V_{f2}$, the critical density is $K_{c2}$, the capacity is $C_2$, the blockage density is $K_{j2}$, and the congestion wave speed is $\omega_2$. Denote $f_{E_{ij},l_{ij}}$ as a mainline cell on the expressways $E_{ij}$, $l_{ij} \in [1, L]$. The on-ramp and off-ramp cells connecting subregion $i$ and expressway $E_{ij}$ are denoted as $f_{iE_{ij}}$ and $f_{E_{ij}i}$, respectively. At time $k$, the flow and density of the mainline cell $f_{E_{ij},l_{ij}}$ are $Q_{E_{ij},l_{ij}}(k)$ and $K_{E_{ij},l_{ij}}(k)$, respectively. For the on-ramp cell $f_{iE_{ij}}$, its flow and density are denoted by $Q_{iE_{ij}}(k)$ and $K_{iE_{ij}}(k)$, respectively, and $\eta_{iE_{ij}}(k)$ denote the on-ramp flow control rate. Similarly, the flow and density of off-ramp $f_{E_{ij}i}$ are denoted by $Q_{E_{ij}i}(k)$ and $K_{E_{ij}i}(k)$, respectively.

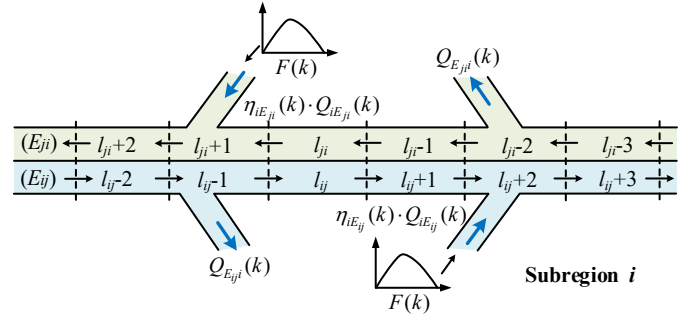

**Fig. 3.** Cell division of expressways.

Vehicles on an expressway and its ramps can be divided into different traffic streams depending on their routes. It is assumed that vehicles with different routes are evenly mixed in the same cell and that the flow transfer satisfies the first-in-first-out principle. Based on this, this study adopts the concept of multi-class CTM to distinguish and track the traffic flow of different routes within the expressway system to realize more accurate traffic flow prediction. Take a cell $f_{E_{ij},l_{ij}}$ as an example, denote the traffic density and flow with route $y \in Y$ within cell $f_{E_{ij},l_{ij}}$ as $K^y_{E_{ij},l_{ij}}(k)$ and $Q^y_{E_{ij},l_{ij}}(k)$. Thus, $K_{E_{ij},l_{ij}}(k) = \sum_{y \in Y} K^y_{E_{ij},l_{ij}}(k)$ and $Q_{E_{ij},l_{ij}}(k) = \sum_{y \in Y} Q^y_{E_{ij},l_{ij}}(k)$.

Based on current traffic density, CTM specifies each cell's traffic demand and receiving capacity. The traffic demand of any route $y \in Y$ in the mainline cell $f_{E_{ij},l_{ij}}$ and on-ramp cell $f_{iE_{ij}}$ are calculated by (1) and (2), respectively.

$$\sigma^y_{E_{ij},l_{ij}}(k) = \min[V_{f1} \cdot K^y_{E_{ij},l_{ij}}(k), C_1] \tag{1}$$

$$\sigma^y_{iE_{ij}}(k) = \min[V_{f2} \cdot K^y_{iE_{ij}}(k), C_2] \tag{2}$$

The receiving capacity of the mainline cell $f_{E_{ij},l_{ij}}$ and on-ramp cell $f_{iE_{ij}}$ are calculated by (3) and (4), respectively.

$$\delta_{E_{ij},l_{ij}}(k) = \min[\omega_1 \cdot (K_{j1} - K_{E_{ij},l_{ij}}(k)), C_1] \tag{3}$$

$$\delta_{iE_{ij}}(k) = \min[\omega_2 \cdot (K_{j2} - K_{iE_{ij}}(k)), C_2] \tag{4}$$

The theoretical flow of a cell is the minimum of the traffic demand and the receiving capacity of its downstream cell. The difference between the inflow and outflow volumes of each cell then results in a change in traffic density. Since the expressway system consists of mainlines, on-ramps, and off-ramps, density dynamics are established separately for the three types of cells.

**a) On-ramps:** The flow of an on-ramp cell $f_{iE_{ij}}$ with route $y$ is:

$$Q_{iE_{ij}}^y(k) = \min[\sigma_{iE_{ij}}^y(k), \frac{\sigma_{iE_{ij}}^y(k)}{\sum_{w \in Y} \sigma_{iE_{ij}}^w(k)} \cdot \delta_{E_{ij},1j}(k)] \qquad (5)$$

where $\sigma_{iE_{ij}}^w(k)$ is the traffic demand of route $w \in Y$ in cell $f_{iE_{ij}}$; cell $f_{E_{ij},1j}$ is the immediate downstream cell of on-ramp cell $f_{iE_{ij}}$ in (5).

The inflow to the on-ramp is generated from the connected subregion, and the flow transmission of the on-ramp is subject to the limitations of its receiving capacity. Assuming that the actual transfer flow from the subregion $i$ to the on-ramp $f_{iE_{ij}}$ with route $y$ is $S_{iE_{ij}}^y(k)$, according to the principle of density conservation, the density of on-ramp cell $f_{iE_{ij}}$ with route $y$ is:

$$K_{iE_{ij}}^y(k+1) = K_{iE_{ij}}^y(k) + [S_{iE_{ij}}^y(k) - \eta_{iE_{ij}}(k) \cdot Q_{iE_{ij}}^y(k)] \cdot T_k / L_s \qquad (6)$$

where $T_k$ (s) is the step length. The outflow rate of on-ramp considers the limitations of ramp metering.

**b) Off-ramps:** Assume that the transfer flow from the expressway $E_{ij}$ to the off-ramp cell $f_{E_{ij}i}$ with route $y$ is $S_{E_{ij}i}^y(k)$. The density of cell $f_{E_{ij}i}$ with route $y$ is:

$$K_{E_{ij}i}^y(k+1) = K_{E_{ij}i}^y(k) + [S_{E_{ij}i}^y(k) - Q_{E_{ij}i}^y(k)] \cdot T_k / L_s \qquad (7)$$

**c) Mainline:** For the mainline cells, if the downstream cell of cell $f_{E_{ij},1j}$ is the merging area, the flow of cell $f_{E_{ij},1j}$ with route $y$ is obtained by (8). If the downstream cell of cell $f_{E_{ij},1j}$ is not the merging area, the $Q_{iE_{ij}}^y(k)$ in (8) is 0.

$$Q_{E_{ij},1j}^y(k) = \min\{\sigma_{E_{ij},1j}^y(k),$$
$$\frac{\sigma_{E_{ij},1j}^y(k)}{\sum_{w \in Y} \sigma_{E_{ij},1j}^w(k)} \cdot [\delta_{E_{ij},1j+1}(k) - \eta_{iE_{ij}}(k) \cdot Q_{iE_{ij}}(k)]\} \qquad (8)$$

The inflow to the mainline comes from its on-ramps and exogenous traffic demand. Suppose that $O_{ij}(k)$ denotes the exogenous traffic demand from subregion $i$ to subregion $j$; $O_{iE_{ij}}(k)$ ($O_{iE_{ji}}(k)$) denotes the traffic demand from subregion $i$ to expressway $E_{ij}$ ($E_{ji}$); $O_{E_{ij}i}(k)$ ($O_{E_{ij}j}(k)$) denotes the traffic demand from expressway $E_{ij}$ to subregion $i$ ($j$), and $O_{E_{ij}E_{ij}}(k)$ denotes the exogenous traffic demand from upstream to downstream of the expressway $E_{ij}$. Newly generated traffic demand upstream of the expressway will enter the first cell. The flow with the route $y$ that can enter the first cell of the expressway $E_{ij}$ is:

$$S_{E_{ij},1j}^y(k) = \min[q^y(k), \frac{q^y(k)}{O_{E_{ij}i}(k) + O_{E_{ij}j}(k) + O_{E_{ij}E_{ij}}(k)} \cdot \delta_{E_{ij},1}(k)] \qquad (9)$$

where $q^y(k)$ is the exogenous traffic demand generated on the route $y$ through traffic assignment, and the assignment method is described in Section IV.

Thus, the density of mainline cell $f_{E_{ij},1j}$ with route $y$ is:

$$K_{E_{ij},1j}^y(k+1) = K_{E_{ij},1j}^y(k) + [Q_{E_{ij},1j-1}^y(k) + \eta_{iE_{ij}}(k) \cdot Q_{iE_{ij}}^y(k) \\ - Q_{E_{ij},1j}^y(k) - S_{E_{ij}i}^y(k)] \cdot T_k / L_s \qquad (10)$$

If cell $f_{E_{ij},1j}$ is not a merging cell, $Q_{iE_{ij}}^y(k)$ is 0; if cell $f_{E_{ij},1j}$ is not a diverging cell, $S_{E_{ij}i}^y(k)$ is 0; if $f_{E_{ij},1j}$ is the first cell, $Q_{E_{ij},1j-1}^y(k)$ will be replaced with $S_{E_{ij},1j}^y(k)$.

Based on the prediction of traffic density, the average speed of each cell is calculated based on the CTM. Taking the mainline cell $f_{E_{ij},1j}$ as an example, the average speed is:

$$V_{E_{ij},1j}(k) = \min\left\{V_{f1}, -\frac{w_1(K_{j1} - K_{E_{ij},1j}(k))}{K_{E_{ij},1j}(k)}\right\} \qquad (11)$$

However, expressway merging zones usually exhibit capacity degradation phenomena [30]. Therefore, this study adopts the method proposed by Han et al. [31] to simulate this phenomenon. The principle is that when the density of a merging cell exceeds a critical value, its maximum outflow volume decreases linearly with its density. If cell $f_{E_{ij},1j}$ is merging bottleneck, the maximum outflow $C_{E_{ij},1j}^{\max}(k)$ of cell $f_{E_{ij},1j}$ is related to the density as:

$$C_{E_{ij},1j}^{\max}(k) = \min\left\{C_b, C_b \cdot \left[1 - \lambda \cdot \frac{K_{E_{ij},1j}(k) - K_{cb}}{K_{j1} - K_{cb}}\right]\right\} \qquad (12)$$

where $C_b$ is the capacity of the merging cell; $K_{cb}$ is the critical density when considering the decrease in the capacity of the merging cell; and $\lambda$ denotes the maximum extent of the capacity drop. If cell $f_{E_{ij},1j}$ is merging bottleneck, $C_1$ in (1) and (3) need to be replaced with $C_{E_{ij},1j}^{\max}(k)$.

### B. MFD-based model for subregion

As a region-based traffic model, MFD relates the average speed to the accumulation of vehicles on a homogeneous road network. This study utilizes MFD for dynamic modeling of the arterial subregion to predict the trip completion rate and accumulation of vehicles in the subregions.

Assume that each subregion within the mixed road network demonstrates a well-defined MFD. Introducing $n_i^y(k)$ to denote the accumulation in subregion $i$ with route $y$ at time $k$, the total accumulation in subregion $i$ is $n_i(k) = \sum_{y \in Y} n_i^y(k)$. As suggested by Hajiahmadi et al. [32], the functional form of the average speed $v_i(k)$ (m/s) of subregion $i$ can be approximated by an exponential function:

$$v_i(n_i(k)) = v_f \exp(-\xi(\frac{n_i(k)}{n_i^{cr}})^\gamma) \tag{13}$$

where $v_f$ (m/s) is the free-flow speed within the subregion and $n_i^{cr}$ (veh) is the critical accumulation of subregion $i$; for a given road network, these values are available. $\xi$ and $\gamma$ are the parameters to be calibrated so (13) can be fitted by actual road network data. The production MFD of homogeneous subregion $i$ is $P_i(n_i(k)) = v_i(n_i(k)) \cdot n_i(k)$.

The trip completion rate for different routes within a subregion is equal to the production ratio to the average trip length. Therefore, the routes within the subregion need to be identified beforehand. Suppose $\alpha$ and $\beta$ denote two random points within a mixed road network. The possible paths $(\alpha, \beta)$ can be: $(i, i)$, two random points within subregion $i$; $(i, \partial_{ij})$, a random point within subregion $i$ and a random point on the boundary between subregion $i$ and subregion $j$; $(i, f_{iE_{ij}})$, a random point within subregion $i$ and a point at the entrance to the on-ramp cell $f_{iE_{ij}}$; $(\partial_{ij}, f_{iE_{ij}})$, a random point on the boundary between subregion $i$ and subregion $j$ and a random point at the entrance to the on-ramp cell $f_{iE_{ij}}$; and $(f_{E_{ji}}, \partial_{ij})$, a random point at off-ramp cell $f_{E_{ji}}$ in subregion $i$ and a random point on the boundary between subregion $i$ and subregion $j$. Denoting the average trip length for travel paths $(\alpha, \beta)$ within subregion as $ATL_{\alpha, \beta}$. For a concrete mixed road network, the trip distance distribution of the above paths can be obtained from experimental data or actual surveys.

Based on the availability of the above path lengths, the trip completion rates for vehicles on different routes within the subregion can be estimated. Outflows from subregions include internal and external transfer flow, whereas external transfer flow includes flows that transfer to adjacent subregions and flows that transfer to connecting expressways. For subregion $i$, the internal trip completion rate is:

$$M_i^y(k) = \frac{n_i^y(k)}{n_i(k)} \cdot \frac{P_i(n_i(k))}{ATL_{i,i}} \quad y = y_{ii} \tag{14}$$

The external trip completion rate of route $y$ from subregion $i$ to adjacent subregion $j$ is:

$$M_i^y(k) = \frac{n_i^y(k)}{n_i(k)} \cdot \frac{P_i(n_i(k))}{ATL_{i,\partial_{ij}}} \quad \forall y \in \{y_{ij}, y_{ijE_{ij}}\} \tag{15}$$

Moreover, the external trip completion rate of route $y$ from subregion $i$ to connecting expressways $E_{ij}$ and expressways $E_{ji}$ are obtained by (16) and (17), respectively:

$$F_i^y(k) = \frac{n_i^y(k)}{n_i(k)} \cdot \frac{P_i(n_i(k))}{ATL_{i,f_{iE_{ij}}}} \quad \forall y \in \{y_{iE_{ij}}, y_{iE_{ij}j}\} \tag{16}$$

$$F_i^y(k) = \frac{n_i^y(k)}{n_i(k)} \cdot \frac{P_i(n_i(k))}{ATL_{i,f_{iE_{ji}}}} \quad y = y_{iE_{ji}} \tag{17}$$

Therefore, the trip completion rate from the subregion to expressways is $F_i(k) = \sum_{y \in Y} F_i^y(k)$, and the trip completion rate only through the arterial network is denoted as $M_i(k) = \sum_{y \in Y} M_i^y(k)$. Finally, according to the principle of accumulation conservation, the traffic dynamics equation for route $y$ in subregion $i$ is:

$$\frac{dn_i^y(k)}{d(k)} = \begin{cases} q^y(k) - M_i^y(k) & y = y_{ii} \\ q^y(k) - \hat{M}_i^y(k) & \forall y \in \{y_{ij}, y_{ijE_{ij}}\} \\ q^y(k) - \hat{F}_i^y(k) & \forall y \in \{y_{iE_{ij}}, y_{iE_{ij}j}, y_{iE_{ji}}\} \end{cases} \tag{18}$$

where $\hat{M}_i^y(k)$ is the transferring flow from subregion $i$ to adjacent subregion with limitation; $\hat{F}_i^y(k)$ is the transferring flow from subregion $i$ to connecting on-ramp with limitation. $\hat{M}_i^y(k)$ and $\hat{F}_i^y(k)$ are obtained by (19) and (20) respectively.

$$\hat{M}_i^y(k) = \min[\mu_{ij}(k) \cdot M_i^y(k),$$
$$\frac{M_i^y(k)}{\sum_{w \in Y} M_i^w(k)} \cdot c_{j,\max} \cdot (1 - \frac{n_j(k)}{n_{j,\max}})] \quad \forall y \in \{y_{ij}, y_{ijE_{ij}}\} \tag{19}$$

$$\hat{F}_i^y(k) = \min[F_i^y(k), \frac{F_i^y(k)}{\sum_{w \in Y} F_i^w(k)} \cdot \delta_{iE_{ij}}(k)] \quad \forall y \in \{y_{iE_{ij}}, y_{iE_{ij}j}\} \tag{20}$$

where $\mu_{ij}(k)$ is the perimeter control rate for transfer demand from subregion $i$ to subregion $j$; $c_{j,\max}$ and $n_{j,\max}$ are the maximum receiving capacity and congestion accumulation of subregion $j$, respectively.

### C. Trip length calibration for a mixed road network

In this section, the SUMO simulation platform calibrates the average trip lengths of different traffic paths within the subregion. This study uses the SUMO to build a mixed road network consisting of an arterial network and expressways, as shown in Fig. 4. The arterial network is connected by two expressways running east-west and north-south, each with two travel directions. The expressway is fully closed and does not connect directly to urban intersections; it can only be connected to the arterial network via on/off ramps. The arterial network in the figure is a 5×5 square grid, with adjacent intersections spaced at 500 m. All sections are three lanes in both directions, with a lane width of 3.5 m. There are 25 intersections in the network, all of which are controlled by signals.

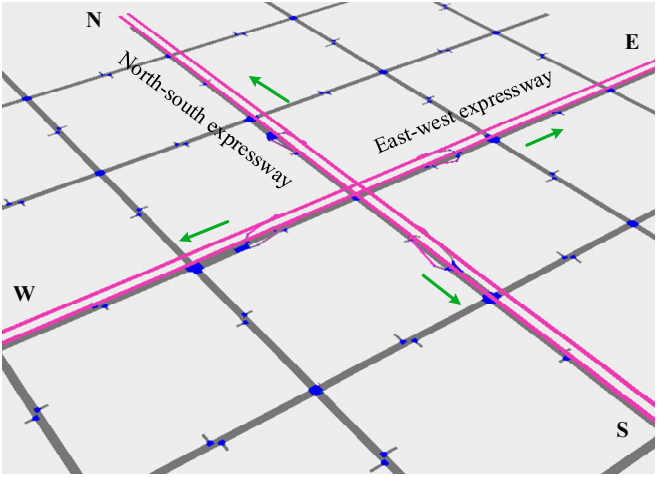

**Fig. 4.** Mixed road network created with SUMO.

On-ramps were provided for each expressway in the network shown in Fig. 4, and traffic detectors were placed in each lane of the on-ramps. In a mixed road network, there are two ways for vehicles to complete their trips: through the arterial network only and from the arterial network to the expressway. The vehicles that complete their trip through only the arterial network are denoted as arterial travel vehicles (ATVs), and the vehicles that complete their trip by entering the expressway via on-ramps are denoted as expressway travel vehicles (ETVs). The proportion of ETV traffic demand to total traffic demand in the whole mixed road network is denoted by $s$. In the experiments, different proportions of arterial network traffic are allocated to the expressway system to study the travel distance distribution of ATVs and ETVs.

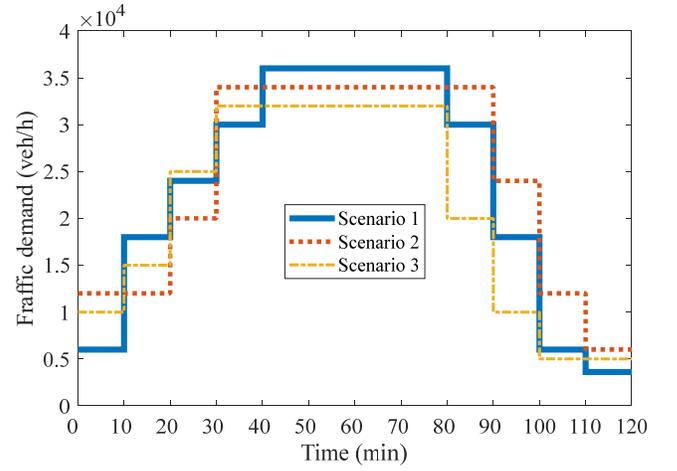

**Fig. 5** Traffic demand loading scenarios.

In SUMO, we use simulated data for traffic loading. In general, we designed three groups of traffic loading scenarios, as shown in Fig. 5, which simulate the process of traffic peak formation and dissipation. Under each group of traffic loading scenarios, the proportion of the traffic demand of the on-ramp to the total traffic demand is then varied separately for the experiments. The total trip completion flow $G$ (veh/s) for the mixed road network consists of two components, i.e., the flow through the on-ramp onto the expressway $F$ (veh/s) and the trip completion flow on the arterial road network $M$ (veh/s). Thus, $G = M + F$. The value of $G$, $F$ and $M$ were obtained by processing the simulation data. Fig. 6 shows the experimental data for $G$, $F$ and $M$ for different values of $s$. The experimental results show that the traffic demand of the expressway for the arterial network changes the shape of the MFD. As $s$ increases, the peak value of $G$ tends to initially rise and then fall, whereas the peak value of $M$ consistently decreases. Meanwhile, The peak value of $F$ gradually increases with increasing $s$. However, the change in $F$ becomes less noticeable when $s$ reaches higher values.

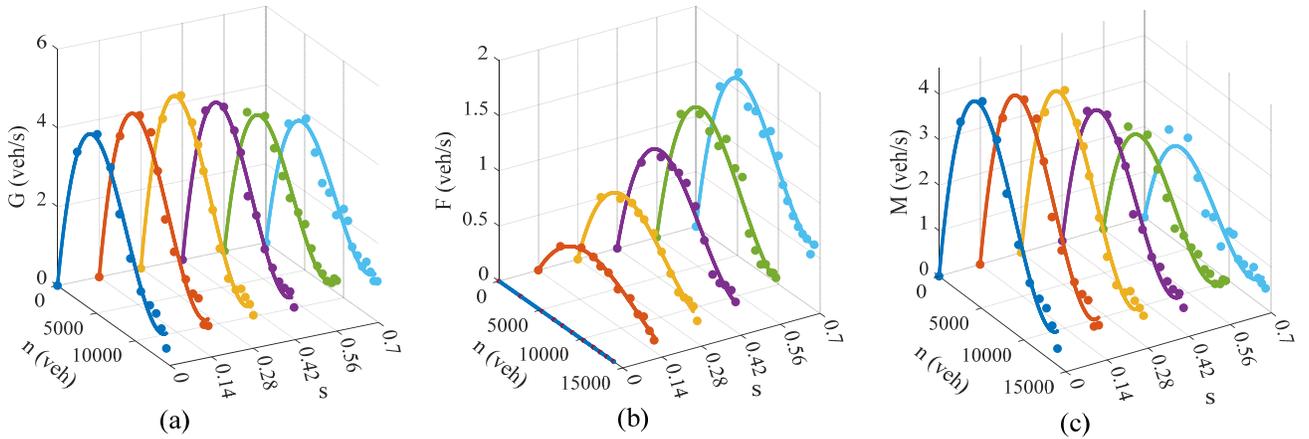

**Fig. 6.** Simulation data for (a) $G$, (b) $F$ and (c) $M$.

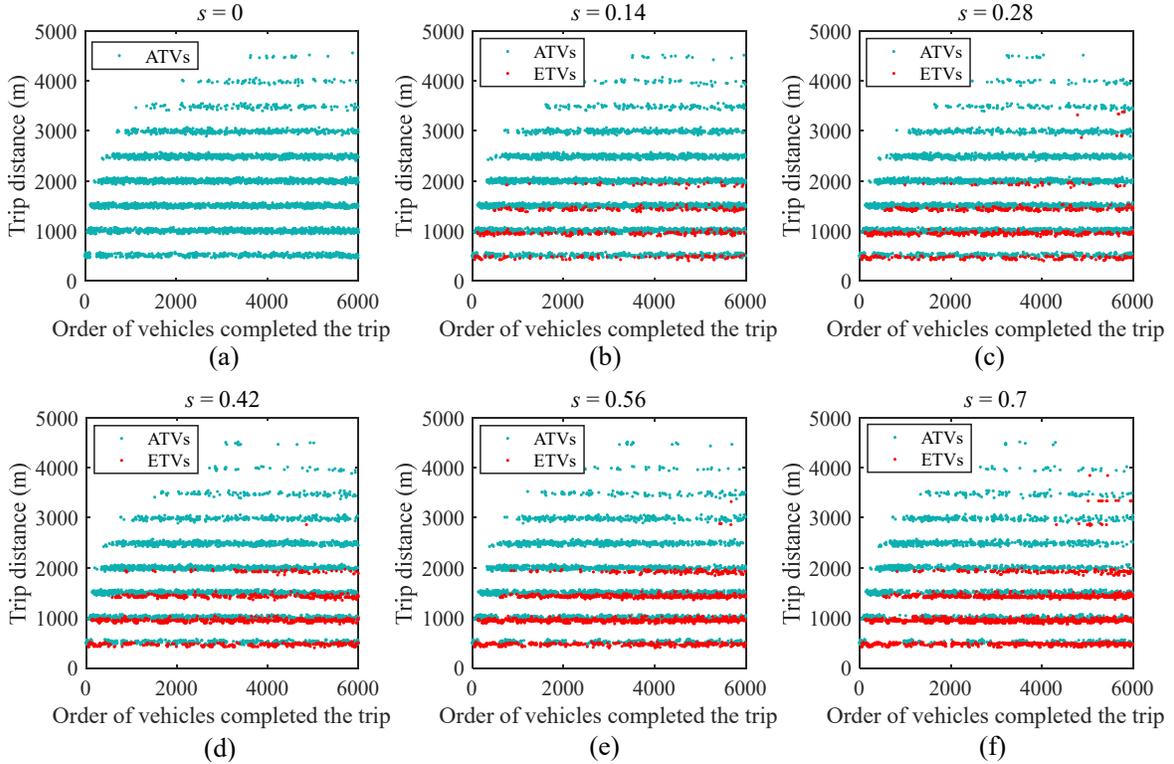

**Fig. 7.** Distribution of vehicle trip lengths within the arterial network.

In each experiment, we considered the trip lengths of the first 6000 vehicles at the end of their trips. Fig. 7 shows the trip length of vehicles; note that only the trip lengths within the arterial network are counted here, i.e., ETVs only count the distance from the origin to the on-ramp. The statistics show that the trip lengths of ATVs are distributed between 500 m and 4500 m, and the trip lengths of ETVs are concentrated between 500 m and 3500 m. The average trip lengths of ATVs and ETVs are 1667 m and 1138 m, respectively. Thus, the total average trip length of vehicles decreases with increasing $s$. Since the average trip length of ETVs is smaller than the trip length of ATVs when the traffic demand of ETVs is small, the total outflow of the network can increase as $s$ increases; however, when the traffic demand of ETVs is significant, the total outflow decreases as $s$ increases due to the limitation of the capacity of the on-ramps. The above results reveal that guiding a portion of the traffic demand to the expressway can improve the efficiency of the road network, and the critical thing is to determine an optimal guidance scheme.

## IV. ROUTE GUIDANCE MODEL FOR MIXED ROAD NETWORKS

The state of the mixed road network is closely related to the route choice method of the vehicles. The average time spent on different travel routes can be estimated by assuming a relatively uniform distribution of vehicles and similar vehicle speeds in a subregion or an expressway cell during the same period. In this section, a route choice model is developed based on the estimation of travel time on different routes, on the basis of which a route guidance strategy is proposed.

The mixed road network has several key points,

including the boundaries of neighboring subregions, entrances to on-ramps, and exits to off-ramps within the subregions. Different travel routes can be divided into segments based on the points mentioned above, and the time spent on the distance belonging to expressways and subregions is estimated, respectively.

For travel paths within the subregion, the average travel time is calculated based on the average trip length and the average speed of the subregion. The average travel time for travel paths $(\alpha, \beta)$ within subregion $i$ can be estimated as:

$$ATT_{\alpha,\beta} = \frac{ATL_{\alpha,\beta}}{v_i(n_i(k))} \tag{21}$$

For travel paths within the expressway system, the average travel time is calculated based on the cell length and the average speed. Regarding ramps, the travel times of the on-ramp cell $f_{iE_{ij}}$ and the off-ramp cell $f_{E_{ij}i}$ are denoted as $TT^r_{iE_{ij}}$ (s) and $TT^r_{E_{ij}i}$ (s), which are calculated through (22) and (23), respectively.

$$TT^r_{iE_{ij}}(k) = \frac{L_s}{V_{iE_{ij}}(k)} \tag{22}$$

$$TT^r_{E_{ij}i}(k) = \frac{L_s}{V_{E_{ij}i}(k)} \tag{23}$$

where $V_{iE_{ij}}(k)$ and $V_{E_{ij}i}(k)$ are average speeds of on-ramp cell

$f_{iE_{ij}}$ and the off-ramp cell $f_{E_{ij}i}$ respectively.

Regarding mainline, the travel time between any two mainline cells $f_{E_{ij},l_a}$ and $f_{E_{ij},l_b}$ ($l_b > l_a$) of expressway $E_{ij}$ is:

$$TT_{l_a,l_b}^e(k) = \frac{(l_b - l_a)L_s}{\sum_{l_{ij}=l_a}^{l_b} V_{E_{ij},l_{ij}}(k)} \qquad (24)$$

TABLE I lists the possible routes between each OD pair in the mixed road network. Based on the origin and destination of the OD, the travel time for each route is calculated as follows:

For vehicles with origin and destination both in subregion $i$, there is only route $y_{ii}$ available. The travel time of the route $y_{ii}$ is:

$$T_{y_{ii}}(k) = ATT_{i,i}(k) \qquad (25)$$

For vehicles that originates in subregion $i$ and terminates in subregion $j$, there are route $y_{ij}$ and route $y_{iE_{ij}j}$ available with respective travel times as:

$$T_{y_{ij}}(k) = ATT_{i,\hat{c}_{ij}}(k) + ATT_{\hat{c}_{ij},j}(k) \qquad (26)$$

$$T_{y_{iE_{ij}j}}(k) = ATT_{i,f_{iE_{ij}}}(k) + TT_{iE_{ij}}^r(k) + TT_{f_{iE_{ij}},f_{E_{ij}j}}^e(k) + TT_{E_{ij}j}^r(k) \qquad (27)$$

For vehicles that originates in subregion $i$ and terminates in expressway $E_{ij}$, there are route $y_{iE_{ij}}$ and route $y_{ijE_{ij}}$ available with respective travel times as:

$$T_{y_{iE_{ij}}}(k) = ATT_{i,f_{iE_{ij}}}(k) + TT_{iE_{ij}}^r(k) + TT_{f_{iE_{ij}},L}^e(k) \qquad (28)$$

$$T_{y_{ijE_{ij}}}(k) = ATT_{i,\hat{c}_{ij}}(k) + ATT_{\hat{c}_{ij},f_{jE_{ij}}}(k) + TT_{jE_{ij}}^r(k) + TT_{f_{jE_{ij}},L}^e(k) \qquad (29)$$

Additionally, for vehicles that originate in expressway $E_{ij}$ and terminate in subregion $j$, there are route $y_{E_{ij}j}$ and route $y_{E_{ij}ij}$ available with respective travel times as:

$$T_{y_{E_{ij}j}}(k) = TT_{1,f_{E_{ij}j}}^e(k) + TT_{E_{ij}j}^r(k) \qquad (30)$$

$$T_{y_{E_{ij}ij}}(k) = TT_{1,f_{E_{ij}i}}^e(k) + TT_{E_{ij}i}^r(k) + ATT_{f_{E_{ij}i},\hat{c}_{ij}}(k) + ATT_{\hat{c}_{ij},j}(k) \qquad (31)$$

Based on the estimation of travel times for the different routes mentioned above, the vehicle spontaneously chooses the appropriate route. When travelers have more than one alternative route, they generally try to choose the route with the shortest travel time. To relax the assumption that travelers are fully informed about traffic, this study uses a logit model to simulate the driver route choice decision without control based on the more realistic concept of stochastic user equilibrium [33]. The idea of the method is to first determine the alternative routes according to the vehicle travel OD, then predict the travel impedance of each alternative route on the arterial network or expressway based on the MFD and CTM, respectively, and finally calculate the selection probability of the alternative route at the current moment using a logit model [34]. Take the traffic demand from subregion $i$ to subregion $j$ as an example, the selection probability of route $y_{ij}$ is:

$$\theta_{y_{ij}}(k) = \frac{\exp(-\mu T_{y_{ij}}(k))}{\exp(-\mu T_{y_{ij}}(k)) + \exp(-\mu T_{y_{iE_{ij}j}}(k))} \qquad (32)$$

where $\theta_{y_{ij}}(k)$ is the drivers' stochastic routing decision based on current network states, and $\mu$ is the logit model parameter that indicates the drivers' knowledge of the mixed road network travel time; thus, a higher $\mu$ corresponds to a higher knowledge of the network travel time. The selection probability of route $y_{iE_{ij}j}$ is $\theta_{y_{iE_{ij}j}}(k) = 1 - \theta_{y_{ij}}(k)$. The route selection probabilities for other traffic demands with more than one alternative route are calculated in the same way.

After modeling vehicle route choice behavior, this study uses a DSO-based route guidance method to influence the traveler's route choice to minimize the total travel time on a mixed road network. However, traffic guidance schemes may not fully meet travelers' own travel convenience needs. When receiving guidance information, it can be assumed that a significant proportion of vehicles will choose the guidance route according to the intentions of the road managers. Assume that the optimized assignment ratio of the route guidance controller on route $y_{ij}$, $y_{iE_{ij}}$ and $y_{E_{ij}j}$ at time $k$ are $\hat{\theta}_{y_{ij}}(k)$, $\hat{\theta}_{y_{iE_{ij}}}(k)$ and $\hat{\theta}_{y_{E_{ij}j}}(k)$ respectively. Thus, the optimized assignment ratio on route $y_{iE_{ij}j}$, $y_{ijE_{ij}}$ and $y_{E_{ij}ij}$ at time $k$ are $1 - \hat{\theta}_{y_{ij}}(k)$, $1 - \hat{\theta}_{y_{iE_{ij}}}(k)$ and $1 - \hat{\theta}_{y_{E_{ij}j}}(k)$ respectively. Then, in actual decision-making, the traveler responds to the route guidance recommendation accordingly and makes the final decision. The actual route choice rate is assumed to be a linear combination of the route guidance recommendation and the driver's route choice rate [35], taking route $y_{ij}$ as an example:

$$\overline{\theta}_{y_{ij}}(k) = \theta_{y_{ij}}(k) + \varepsilon(\hat{\theta}_{y_{ij}}(k) - \theta_{y_{ij}}(k)) \qquad (33)$$

where $\overline{\theta}_{y_{ij}}(k)$ is the actual route choice rate of travelers and $\varepsilon$ is the assumed compliance rate with the optimized route guidance, indicating the proportion of drivers following the route guidance recommendations.

To model the state of the road network dynamically, the generated traffic demand $O_{ij}(k)$ at time $k$ is split according to the route. Denote by $q^{y_{ij}}(k)$ and $q^{y_{iE_{ij}j}}(k)$ the actual traffic demand assigned to route $y_{ij}$ and route $y_{iE_{ij}j}$, respectively, $O_{ij}(k) = q^{y_{ij}}(k) + q^{y_{iE_{ij}j}}(k)$ . The traffic demand generated on route $y_{ij}$ at moment $k$ under the guidance scheme can be obtained:

$$q_{y_{ij}}(k) = \overline{\theta}_{y_{ij}}(k) \cdot O_{ij}(k) \tag{34}$$

The traffic demand for other OD pairs in TABLE I with more than one alternative route is assigned in the same way.

## V. COOPERATIVE GUIDANCE AND CONTROL METHODS FOR MIXED ROAD NETWORKS

### A. Control Objectives

The mixed road network traffic model developed in this study provides a prediction method for the traffic state of expressways and subregions. Based on this, this study employs route guidance and flow control measures in the strategy formulation module to solve the road network congestion problem. Since the results of guidance and control are mutually influential, separate decisions may cause misjudgments, resulting in over- or underregulation. Therefore, this study integrates road guidance strategies with perimeter control and ramp metering to develop a synergistic optimization model for traffic demand management and traffic system control on mixed road networks.

For mixed road networks, cooperative guidance and control aim to improve overall traffic efficiency and reduce congestion. The TTS of all the vehicles traveling on the network is usually used as an indicator of traffic efficiency and is adopted as the control objective, denoted as:

$$J = \min$$

$$T_k \cdot \sum_{k=0}^{K-1} \left( \sum_{i=1}^{2} n_i(k) + L_s \cdot \sum_{i,j \in [1,2], i \neq j} \left( K_{iE_{ij}}(k) + \sum_{l_{ij}=1}^{L} K_{E_{ij},l_{ij}}(k) \right) \right) \tag{35}$$

subject to:

$$0 \leq n_i(k) \leq n_{i\max} \quad i = 1,2 \tag{36}$$

$$0 \leq u_{ij}(k) \leq 1 \quad i = 1,2; j = 3 - i \tag{37}$$

$$0 \leq \eta_{iE_{ij}}(k), \eta_{iE_{ji}}(k) \leq 1 \quad i,j = 1,2 \tag{38}$$

$$0 \leq \hat{\theta}_{y_{ij}}(k), \hat{\theta}_{y_{iE_{ij}}}(k), \hat{\theta}_{y_{E_{ij}j}}(k) \leq 1 \quad i \neq j \tag{39}$$

and (1)-(34).

The guidance and control decisions under the cooperative guidance and control (CGC) policy are made considering the interaction between them. Thus, the premise of constructing the cooperative strategy is the real-time connection of traffic guidance information and traffic control information. The two tasks will be performed separately if guidance and control are not well connected. Noncooperative guidance and control (NCGC) is a hierarchical decision-making process in which the guidance variable is obtained by solving the current traffic state and demand with $J_G$ as the optimization objective.

$$J_G = J$$
$$u_{ij}(k) = 1 \tag{40}$$
$$\eta_{iE_{ij}}(k), \eta_{iE_{ji}}(k) = 1$$

When making control decisions, since the traffic guidance scheme is unknown, the outcome of the traffic assignment is predicted based only on the logit model. Thus, the optimization objective is:

$$J_C = J$$
$$\overline{\theta}_{y_{ij}}(k) = \theta_{y_{ij}}(k)$$
$$\hat{\theta}_{y_{iE_{ij}}}(k) = \theta_{y_{iE_{ij}}}(k) \tag{41}$$
$$\hat{\theta}_{y_{E_{ij}j}}(k) = \theta_{y_{E_{ij}j}}(k)$$

CGC is a cooperative control method for perimeter control and route guidance that achieves interaction between the two systems through real-time information sharing. In CGC, the flow control and route guidance systems can understand each other's decisions and adjust accordingly to optimize the state of the overall road network. In contrast, NCGC regards flow control and route guidance as independent systems lacking information sharing and cooperative decision-making.

### B. Solving Methods

The mixed road network traffic model can predict flows, and this study builds a multi-input, multi-output control model with constraints. Therefore, it is suitable for MPC to solve the optimal control problem. MPC can use feedback monitoring information for a dynamic system to address the gap between the desired and actual values. MPC evaluates the real-time system state and prediction model data using an objective function in a finite horizon at each controller sampling moment. The MPC repeats the optimization process of the previous moment at each controller sampling moment, solving a new finite-horizon optimization problem based on the new state information to achieve rolling optimization.

The key step in utilizing MPC is to solve for the optimal control variables based on the objective. This study uses the particle swarm optimization (PSO) algorithm. PSO effectively handles complex nonlinear problems, multiobjective optimization, and global optimum search in MPC. PSO possesses the capability of global exploration and can simultaneously process multiple solutions. Therefore, integrating PSO with MPC overcomes the limitations of traditional optimization methods, providing a robust solution approach for MPC. Suppose the prediction horizon is $N_p$, the control horizon is $N_c$, the time sequence of the controller is $k_c$, the step length of the controller is $T_c$, and the control scheme at $k_c$ is $U(k_c)$ (including the guidance schemes $\hat{\theta}_{y_{ij}}(k_c)$, $\hat{\theta}_{y_{iE_{ij}}}(k_c)$ and $\hat{\theta}_{y_{E_{ij}j}}(k_c)$, and the control schemes $u_{ij}(k_c)$, $\eta_{iE_{ij}}(k_c)$ and $\eta_{iE_{ji}}(k_c)$). The CGC process for the mixed road network is shown in Fig. 8. The above process can also be used to solve the NCGC problem; the difference is that when solving the NCGC problem, (42) and (43) are used as the objective functions for prediction and solution, respectively.

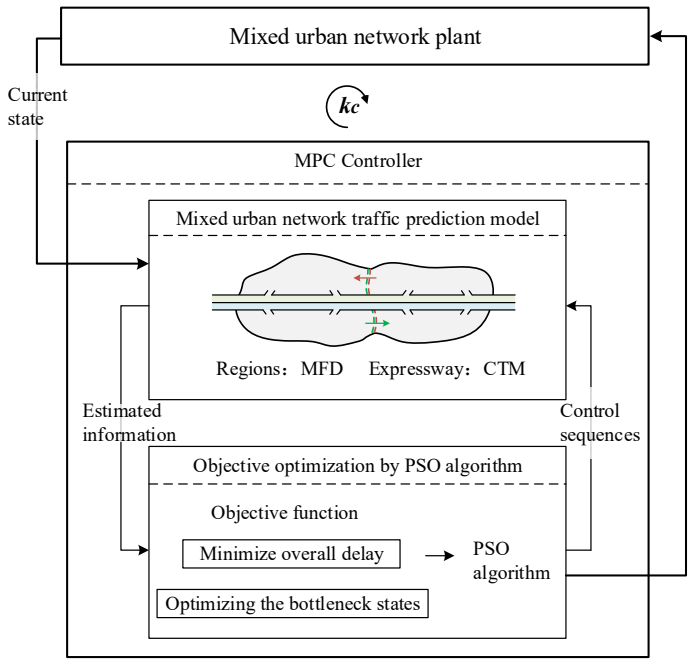

**Fig. 8.** MPC control flow chart.

## VI. CASE STUDIES

### A. Scenario Description

In this section, we analyze the effectiveness of the proposed scheme through a case study. The mixed road network structure shown in Fig. 1 is used in the experiments. The total number of cells in each direction of the expressway is 17, and the length of each cell is $L_s$ =500 m. The serial numbers of merging cells are $f_{1E_{12}} = 7$, $f_{1E_{21}} = 15$, $f_{2E_{12}} = 15$ and $f_{2E_{21}} = 7$, and the serial numbers of diverging cells are $f_{E_{12}1} = 3$, $f_{E_{12}2} = 11$, $f_{E_{21}1} = 11$ and $f_{E_{21}2} = 3$. The free-flow speed of the expressway mainline is $V_{f1}$ =80 km/h, and the capacity is $C_1$ =6000 veh/h; the free-flow speed of the ramps is $V_{f2}$ =40 km/h, and the capacity is $C_2$ =3000 veh/h. The capacity of the merging bottleneck is set to $C_b$ =4800 veh/h, and the capacity decrease parameter $\lambda$ =0.3. For simplicity, it is assumed that the two subregions are identical and that both adopt the road network used for the experiments in Section III. The critical accumulation in the subregion is $n_i^{cr}$ =4650 veh, and the blocking accumulation in the subregion is $n_{i\max}$ =13000 veh. The experimental data show that the free-flow speed of the adopted road network is approximately 9 m/s, and fitting the speed function of (15) results in $\varsigma$ =1.286 and $\gamma$ =1. The initial conditions are set to $n_1(0)$ =6000 veh and $n_2(0)$ =3000 veh. The simulation step length in the experiments is $T_s$ =10 s, and the control step length of the controller is $T_c$ =60 s. The parameters $\mu$ and $\varepsilon$ are to be determined and will be discussed in the experiments. The trapezoidal traffic demand shown in Fig. 9 is used for the mixed road network in the numerical simulation.

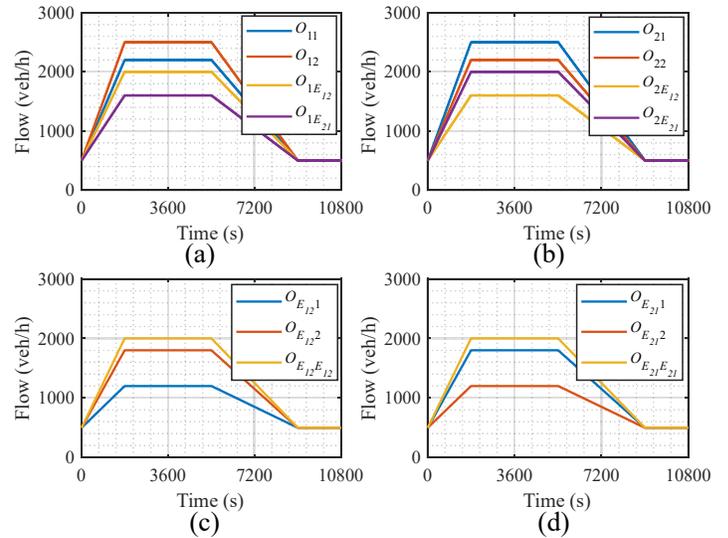

**Fig. 9.** Traffic demand for the mixed road network.

To analyze and compare the impact of the synergy of traffic guidance and control on the overall traffic of the mixed road network, the following policies are implemented under the same conditions: no guidance and no control (NGNC); noncooperative guidance and control (NCGC); and cooperative guidance and control (CGC).

### B. Result Analysis

For simplicity, subregion 1 and subregion 2 are named as SR1 and SR2, respectively. The sensitivity parameter $\mu$ describes the ability of the travel group to perceive traffic information. The simulation results for the NGNC strategy with $\mu$ =0.5 are shown in Fig. 10. Figs. 10(a) and (d) indicate that the merging bottlenecks all experience significant congestion after peak arrival, which lasts for a long time. Figs. 10(b) and (e) show the trend of the accumulations for SR1 and SR2, with average values of 5354 veh and 4688 veh, respectively, during the simulation period. Figs. 10(c) and (f) show the trip completion flow for SR1 and SR2, with average values of 3.676 veh/s and 3.656 veh/s, respectively.

The accumulations of the expressways and the subregions for different values of $\mu$ under the NGNC policy are shown in Fig. 11. The mean total accumulations of the mixed road network are 11319 veh, 11150 veh, and 10901 veh for $\mu$ =0.05, 0.5, and 2, respectively. Thus, the network status shows an improving trend with increasing $\mu$; however, the difference is not significant. The expressway status for different values of $\mu$ is almost the same, while the states of SR1 and SR2 are more similar when $\mu$ is larger. The results show that as the ability of the traveler group to perceive the network state increases, the overall state of the mixed road network becomes more balanced.

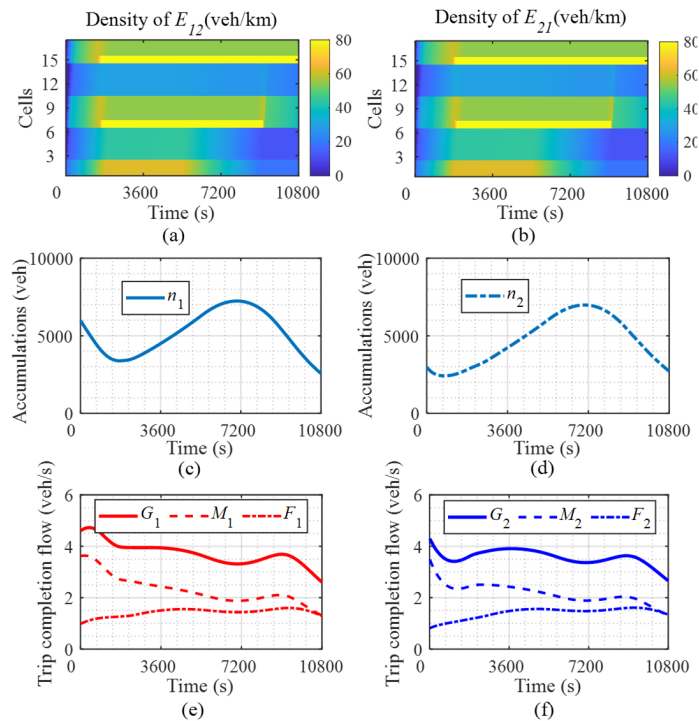

**Fig. 10.** Simulation results for the NGNC strategy with $\mu = 0.5$.

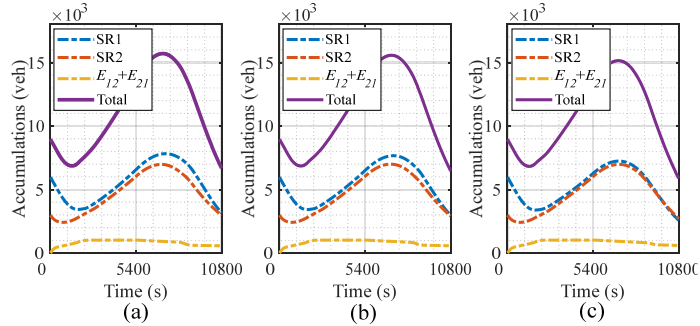

**Fig. 11.** Accumulations for the NGNC policy with $\mu$ taking values of (a) 0.05, (b) 0.5 and (c) 2.

When employing the NCGC policy, in addition to the sensitivity parameter, the parameter of traveler compliance with the guidance signal, $\varepsilon$, affects the control results. Taking $\mu = 0.5$ and $\varepsilon = 0.5$ as an example, Fig. 12 shows the simulation results with the NCGC strategy. There is consistently no significant congestion on the expressways. The average accumulations of SR1 and SR2 are 4924 veh and 4132 veh, which are 8.03% and 11.86% lower than those of the NGNC policy. Thus, regional congestion is relieved under the NCGC strategy compared to the NGNC strategy. The average trip completion flows of SR1 and SR2 are 3.958 veh/s and 3.978 veh/s, respectively, and the results are not significantly different from those of the NGNC policy.

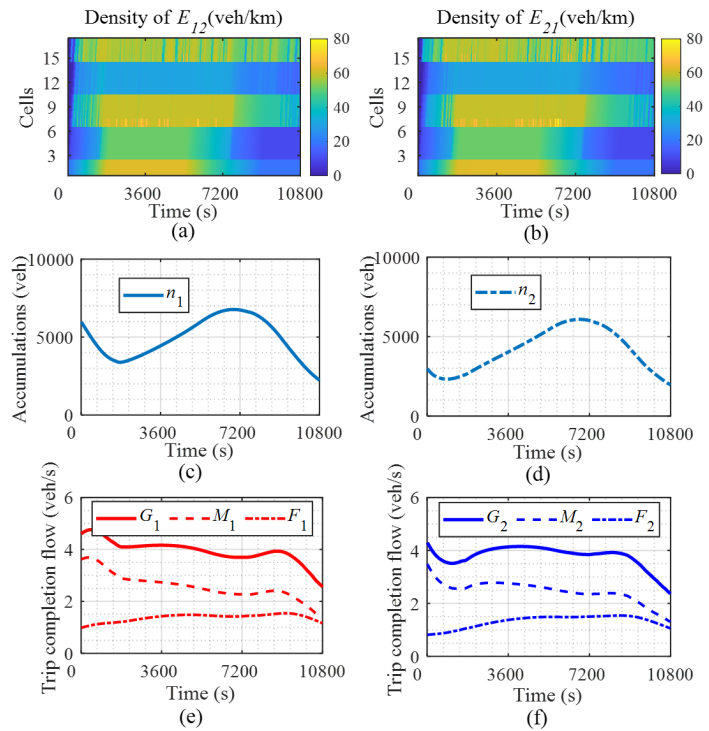

**Fig. 12.** Simulation results under the NCGC policy.

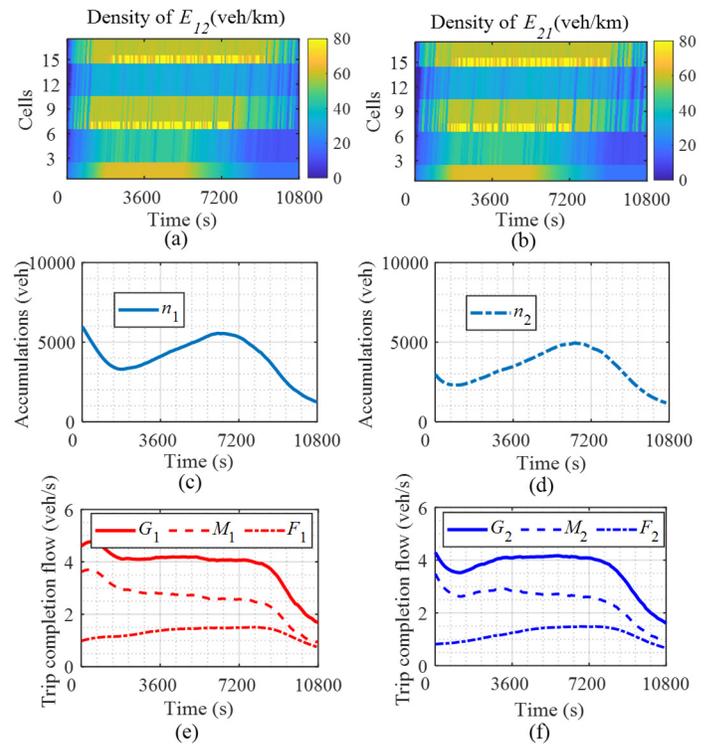

**Fig. 13.** Simulation results under the CGC policy.

For the CGC policy, the same experiment was conducted with $\mu = 0.5$ and $\varepsilon = 0.5$, and the simulation results are shown in Fig. 13. Benefiting from traffic guidance and control, the traffic conditions in each expressway are maintained almost optimally during the peak hours (Figs. 13 (a) and (d)). Over the simulation period, the average accumulations in SR1 and SR2 are 3930 veh and 3351 veh, which are 20.19% and 18.90% lower than those of the NCGC policy. The average trip completion flows in SR1 and SR2 are 4.073 veh/s and 3.990 veh/s, respectively. Therefore, compared with the NGNC and NCGC strategies, the CGC strategy improves traffic efficiency while reducing congestion in the network.

Figs. 12 and 13 visualize the phenomenon whereby the trip completion flow can remain high during the peak period as the accumulations on the mixed road network gradually increase. This phenomenon reflects that the MFD of the mixed road network is related to the accumulations and the ETV traffic demand. Figs. 14(a) and (d) show the trend of outflow for SR1 and SR2 under the NGNC, NCGC, and CGC policies (the black dot indicates the origin). Figs. 14(b)-(c) and 14(e)-(f) show the side and top views of Figs. 14(a) and (d), respectively. The guidance decision influences the trip completion flow of the mixed road network, as it can continuously regulate the traffic demand flowing to the expressway.

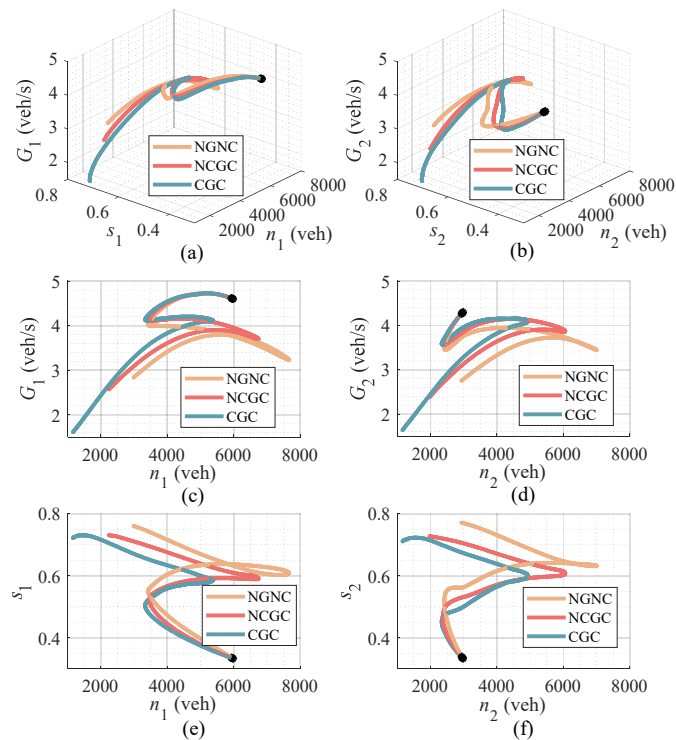

**Fig. 14.** Trip completion flows of mixed road networks under the NGNC, NCGC, and CGC policies.

To compare the control effectiveness of the three policies, the total trip completion flow and the total accumulations of the mixed road network are quantified. For the expressway, the flows of the last mainline cells are taken as the trip completion flows. Therefore, for each control step, the trip completion flow of the expressways is $Q_{E_{12},L}(k) + Q_{E_{21},L}(k)$. A comparison of total trip completion flow and total accumulations under the NGNC, NCGC, and CGC policies is shown in Fig. 15. The CGC strategy has the highest traffic efficiency and the least congestion during peak periods. The statistical results of the average total trip completion flow and average total accumulations under the three policies are shown in TABLE II and TABLE III. The average total trip completion flow for the

CGC policy during the peak and congestion relief period (1800 s~9000 s) is 8.9% higher than that of the NGNC policy and slightly higher than that of the NCGC policy. The average total accumulation for the CGC policy is 25.99% and 11.81% lower than that of the NGNC and NCGC policies.

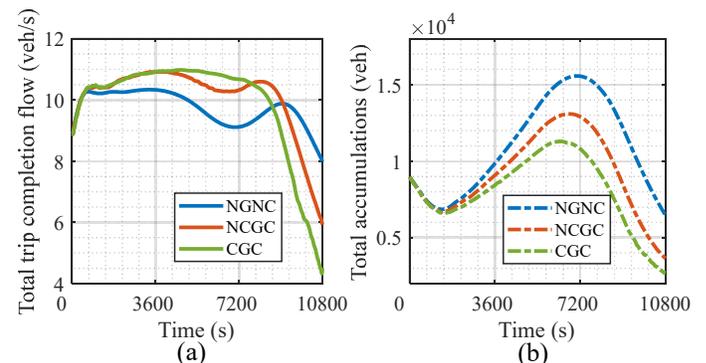

**Fig. 15.** (a) Total trip completion flow and (b) total accumulations under the NGNC, NCGC and CGC policies.

TABLE II
AVERAGE TRIP COMPLETION FLOWS (veh/s)

|      | SR1   | SR2   | $E_{12}$ | $E_{21}$ | Total  |
|------|-------|-------|----------|----------|--------|
| NGNC | 3.676 | 3.656 | 1.213    | 1.216    | 9.761  |
| NCGC | 3.958 | 3.978 | 1.220    | 1.215    | 10.371 |
| CGC  | 4.073 | 3.990 | 1.307    | 1.311    | 10.681 |

TABLE III
AVERAGE ACCUMULATIONS (veh)

|      | SR1  | SR2  | $E_{12}$ | $E_{21}$ | Total |
|------|------|------|----------|----------|-------|
| NGNC | 5354 | 4688 | 560      | 548      | 11150 |
| NCGC | 4924 | 4132 | 475      | 480      | 10011 |
| CGC  | 3930 | 3351 | 494      | 484      | 8259  |

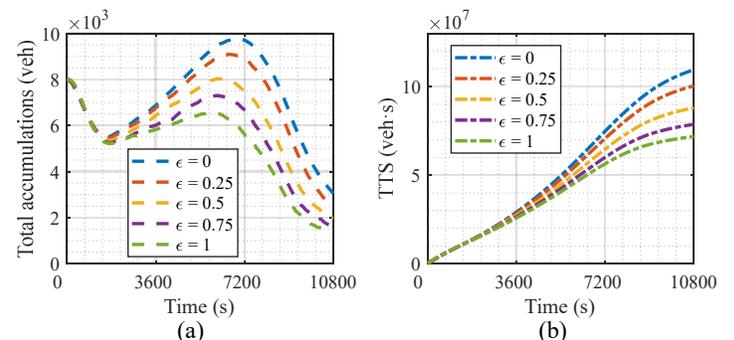

**Fig. 16.** Simulation results of the CGC policy for different values of $\varepsilon$.

The effectiveness of the guidance strategy is determined by the parameter $\varepsilon$ of driver compliance with guidance signals. When $\varepsilon$ is 0, the traffic guidance function is deficient, and only traffic control works; when $\varepsilon$ is 1, all drivers follow the guidance signals, which may only be possible in special scenarios, such as fully connected and autonomous driving environments. In the CGC policy, simulations were performed with $\varepsilon$ set to 0, 0.25, 0.5, 0.75, and 1. The results in Fig. 16 show that with increasing $\varepsilon$, the total accumulation and TTS within the mixed road network gradually decrease. This result indicates that effective implementation of the guidance strategy can improve traffic efficiency and reduce traffic congestion. The control effect is weakest when $\varepsilon$ is 0, proving that cooperative traffic guidance and control is more advantageous than using traffic control measures only.

## VII. CONCLUSION

This study aims to establish the traffic transmission mechanism of mixed road networks consisting of expressways and arterial networks and utilizes macroscopic traffic strategies to address congestion issues within mixed road networks. The main work is divided into three parts. Firstly, integrating the multi-class CTM with the MFD model achieves effective tracking of vehicle paths within expressway systems and arterial subregions, thus integrating the traffic dynamics of the mixed road network. Secondly, it calibrates the average trip length of expressway traffic flow using experimental data. Finally, an integrated control method combining route guidance, perimeter control, and ramp metering is proposed, and case studies demonstrate the superiority of the cooperative traffic guidance and control strategy. Two conclusions are obtained from this study. First, in the mixed road network, the peak of the total travel completion flow tends to increase and decrease as the on-ramp traffic demand increases. Second, the case study results demonstrate that the CGC policy can improve the total traffic efficiency of the mixed road network, and the proposed integrated optimization objective can reduce the risk of congestion propagation at expressway bottlenecks while alleviating the overall congestion of the mixed road network.

Expressways may take on different levels of transport pressure for trips between subregions across different distances. Therefore, the relationship between the attractiveness of expressways for trips and the distance between subregions deserves further study, and the corresponding control schemes must be further adjusted. In addition, with the rise of connected and autonomous driving technology, there may be a mixture of human-driven vehicles (HVs) and connected and autonomous vehicles (CAVs) in urban networks. Work has been conducted to confirm the existence of MFD in mixed driving environments [36]. This study has demonstrated that the driver population's level of compliance with guidance information is one of the most critical factors affecting the efficiency of cooperative control. Increased CAV market penetration may promote the controllability of road network systems; therefore, control measures in such environments deserve further study.

## ACKNOWLEDGMENT

This research was financially supported by the National Natural Science Foundation of China (Grant Nos. 52072108 and 52372326), the Municipal Natural Science Foundation of Hefei (Grant No. 2022020) and China Scholarship Council awards.

## About the Authors

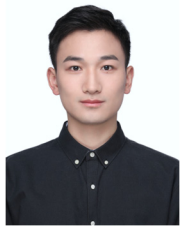
**Yunran Di** received a B.Eng. degree from the North China University of Technology, Beijing, China, in 2014; an MS degree from the School of Automotive and Transportation Engineering at Hefei University of Technology, Hefei, Anhui Province, China, in 2018. He is currently pursuing his Ph.D. degree in engineering at Hefei University of Technology. His research interests include connected autonomous driving, regional road network traffic modeling, and traffic control. Now he is a visiting scholar at the University of Wisconsin-Madison, USA.

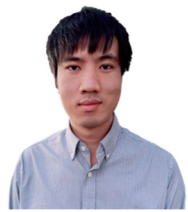
**Haotian Shi** serves as a research associate at the University of Wisconsin-Madison. He received his Ph.D. degree in Civil and Environmental Engineering from the University of Wisconsin-Madison in May 2023. He also received three MS degrees in Power and Machinery Engineering (Tianjin University, 2020), Civil and Environmental Engineering (UW-Madison, 2020), and Computer Sciences (UW-Madison, 2022). His main research directions are prediction/control of connected automated vehicles, intelligent transportation systems, traffic crash data analysis, and deep reinforcement learning.

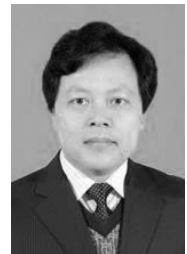
**Weihua Zhang** received a B.Eng. degree from the Automobile Department, Chang'an University, Xi'an, Shanxi, China in 1991; an MS degree from the Mechanics Institute, Hefei University of Technology, Hefei, Anhui, China in 1998; and a DE degree from the School of Transportation, Southeast University, Nanjing, Jiangsu, China in 2003. His research interests include the theory and method of regional and urban transportation system planning, the basic theory and method of intelligent transportation system, traffic environment, energy consumption, and traffic safety. He has had more than 70 academic papers published in core or important academic journals and academic conferences.

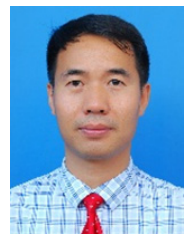
**Heng Ding** received his B.Eng. and MS degrees from the School of Information Engineering at Chang'an University, Xi'an, Shanxi Province, China, in 2002 and 2005 and his DE degree from the School of Transportation Engineering, Hefei University of Technology, Hefei, Anhui province, China in 2011. He joined the Department of Traffic Engineering of Hefei University of Technology (now School of Automotive and Transportation Engineering, Hefei University of Technology) and has engaged in scientific research and teaching. Mr. Ding is currently a Professor with the School of Automotive and Transportation Engineering, a branch secretary of the Department of Road and Traffic Engineering and a deputy director of the Institute of Traffic Engineering, Hefei University of Technology, Hefei. His research interests include traffic management and control, intelligent measurement, and control technology and application.

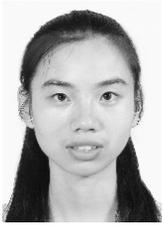

**Xiaoyan Zheng** received her B.Eng., MS and DE degrees from the School of Highway Engineering, Chang'an University, Xi'an, Shanxi, China in 2002, 2005 and 2018. She is currently with the School of Automotive and Transportation Engineering, Hefei University of Technology, Hefei, China. She worked at the Hefei University of Technology after graduating from the Department of Traffic Engineering. She started focusing on bridge and tunnel engineering when she pursued her doctoral degree at Chang'an University in 2008. Her research interests include bridge durability, bridge performance design methods, and traffic flow analysis theory and methods.

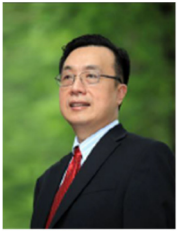

**Bin Ran** is the Vilas Distinguished Achievement Professor and Director of ITS Program at the University of Wisconsin at Madison. Dr. Ran is an expert in dynamic transportation network models, traffic simulation and control, traffic information system, Internet of Mobility, Connected Automated Vehicle Highway (CAVH) System. He has led the development and deployment of various traffic information systems and the demonstration of CAVH systems. Dr. Ran is the author of two leading textbooks on dynamic traffic networks. He has co-authored more than 240 journal papers and more than 260 referenced papers at national and international conferences. He holds more than 20 patents of CAVH in the US and other countries. He is an associate editor of Journal of Intelligent Transportation Systems.